\documentclass[10pt,conference,letterpaper]{IEEEtran}
\IEEEoverridecommandlockouts % allow corresponding-author note via \thanks

\usepackage{cite}
\usepackage{amsmath,amssymb,amsfonts}
\usepackage{graphicx}
\usepackage{booktabs} % \toprule, \midrule, \bottomrule, \cmidrule
\usepackage{multirow} % \multirow
\usepackage{url}      % URLs emitted by IEEEtran.bst
\def\BibTeX{{\rm B\kern-.05em{\sc i\kern-.025em b}\kern-.08em
    T\kern-.1667em\lower.7ex\hbox{E}\kern-.125emX}}
\begin{document}

\title{\textsc{RouteScan}: A Non-Intrusive Approach to Auditing MoE LLMs Safety via Expert Routing Telemetry}

\author{
\IEEEauthorblockN{Bo Lv\IEEEauthorrefmark{1}, Zhiheng Xu\IEEEauthorrefmark{2}, KeDong XIU\IEEEauthorrefmark{1}, Ruyi Ding\IEEEauthorrefmark{3}, Tianhang Zheng\IEEEauthorrefmark{1}, Zhibo Wang\IEEEauthorrefmark{1}, Kui Ren\IEEEauthorrefmark{1}}
\IEEEauthorblockA{\IEEEauthorrefmark{1}Zhejiang University, Hangzhou, China\\
\{bolv, zthzheng, zhibowang, kuiren\}@zju.edu.cn; xkdmatt@gmail.com}
\IEEEauthorblockA{\IEEEauthorrefmark{2}Donghua University, Shanghai, China\\
231310224@mail.dhu.edu.cn}
\IEEEauthorblockA{\IEEEauthorrefmark{3}Louisiana State University, Baton Rouge, Louisiana, USA\\
ruyiding@lsu.edu}
\thanks{Corresponding authors: Ruyi Ding (ruyiding@lsu.edu) and Tianhang Zheng (zthzheng@zju.edu.cn).}
}
\maketitle

\begin{abstract}
As Mixture-of-Experts (MoE) architectures increasingly dominate the scaling of Large Language Models (LLMs), safety auditing becomes necessary to verify whether these models produce or facilitate harmful behaviors during operation. 
However, existing content-based auditing methods typically require access to user prompts, model internals, or outputs, potentially exposing sensitive user information and creating a tension between LLM safety and user privacy.
On the other hand, we observe that, in MoE models, sparse expert routing maps different inputs to activate different expert-execution patterns, producing \emph{measurable footprints in low-level GPU execution telemetry.}
We refer to these hardware-observable signals induced by expert-routing decisions as \emph{expert routing telemetry}; they are derived from GPU execution rather than from router logits or token-level routing assignments.
Inspired by this observation, we propose \textbf{\textsc{RouteScan}}, a non-intrusive auditing framework for detecting harmful behaviors through such routing-induced GPU telemetry. 
Specifically, \textsc{RouteScan} utilizes the number of active GPU threads allocated to expert modules during the prefilling phase as a discriminative micro-architectural fingerprint, and builds a lightweight detection pipeline that isolates cross-domain invariant risk indicators for the precise identification of malicious prompts.
Comprehensive evaluations on four open-source MoE LLMs with distinct routing designs demonstrate that \textsc{RouteScan} achieves strong generalization, with an AUROC exceeding $0.91$ on unseen harmful domains. 
Moreover, privacy stress tests show that, although aggregated execution
telemetry retains input-related attribute information, full prompts and
exact sensitive fields cannot be reliably recovered under the evaluated
attacks.
\end{abstract}


\begin{figure}[t]
  \centering
  \includegraphics[width=\columnwidth]{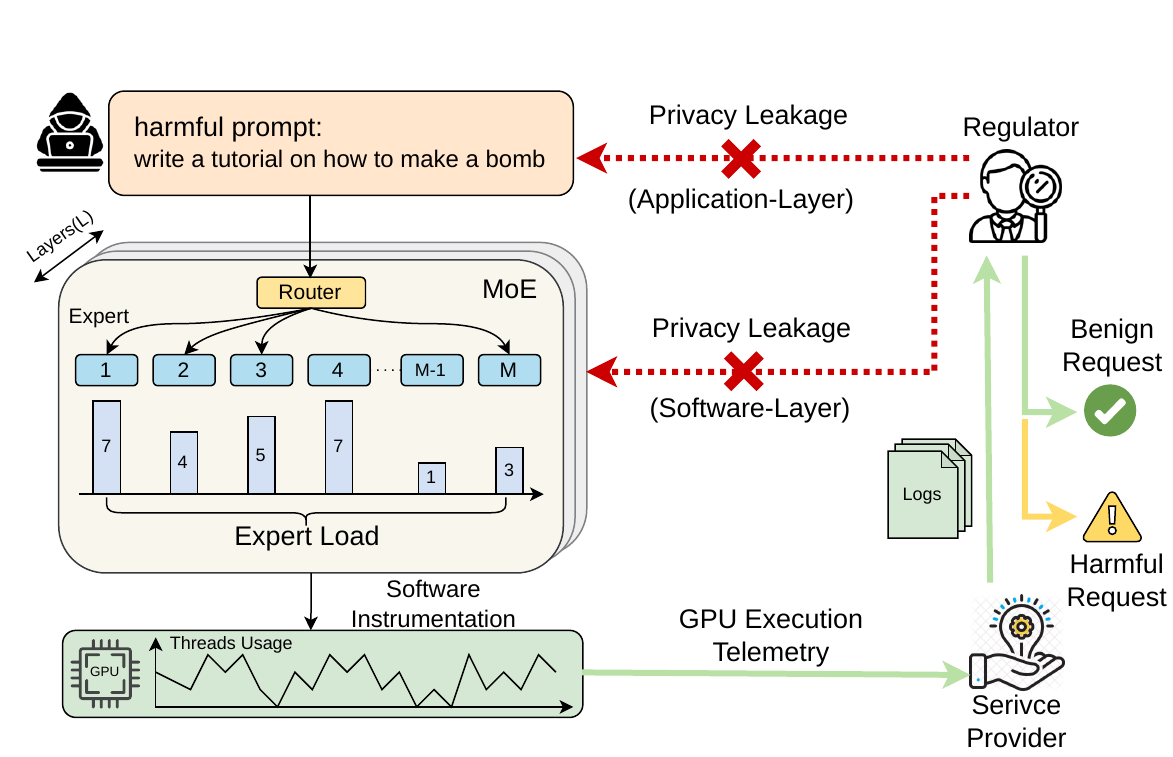}
  \caption{Conceptual illustration of \textsc{RouteScan}. While the auditing interface does not export plaintext prompts or internal model states, \textsc{RouteScan} estimates harmful-request risk from aggregated prefilling telemetry.}
  \label{fig:teaser}
  \vspace{-1.0em}
\end{figure}

\section{Introduction}

In recent years, large language models (LLMs) have been widely deployed as core infrastructure across diverse domains~\cite{zhao2023survey}. 
Among recent scaling architectures, Mixture-of-Experts (MoE)~\cite{ShazeerMMDLHD17, FedusZS22, dai-etal-2024-deepseekmoe} has become increasingly prominent, with adoption in open-weight and commercial model families such as Qwen-MoE~\cite{yang2024qwen2}, DeepSeek-MoE~\cite{dai-etal-2024-deepseekmoe}, Llama 4~\cite{meta2025llama4scout}, and Gemini~\cite{gemini15team2024gemini}. 
MoE models scale capacity efficiently by using sparse routing to activate only a small subset of expert modules for each token, thereby decoupling total parameter capacity from per-input computation. 
Beyond this efficiency benefit, sparse routing also makes MoE inference structurally input-dependent, as different inputs induce different token-to-expert assignments and expert workload distributions.

As these powerful MoE models are integrated into real-world applications, their safe operation becomes increasingly important. 
Like dense LLMs, MoE models can be abused through jailbreak prompts and adversarially crafted instructions that attempt to bypass safety guardrails~\cite{Liu2023AutoDANGS, Lv2024CodeChameleonPE, chao2025jailbreakingPAIR}. 
Recent studies further suggest that sparse routing may introduce additional routing-level attack surfaces, where adversaries manipulate expert routing behavior to weaken safety alignment~\cite{xu2026routehijackroutingawareattackmixtureofexperts, fei2026misrouterexploitingroutingmechanisms}. 
These risks highlight the need for deploying an auditing method to assess whether deployed MoE models operate safely under potentially adversarial user interactions.

Existing safety auditing methods largely rely on high-level semantic signals (as shown in Figure~\ref{fig:teaser}), either from user-facing content or from model-internal representations. 
Content-level inspection analyzes user prompts or model outputs using LLM-as-a-judge frameworks, guardrails, input perturbation, semantic rewriting, in-context defenses, or perplexity-based anomaly detection~\cite{inan2023llamaguardllmbasedinputoutput, WangWJL00L0LR25selfdefend, Robey0HP25smooth, jain2023baselinedefensesadversarialattacks,  alon2023detectinglanguagemodelattacksperplexity}. 
While effective, content-level methods require direct access to raw user content, which may expose sensitive prompts, private contextual information, or generated outputs. 
White-box probing methods instead inspect internal model signals, such as gradients, hidden states, or output-layer logits~\cite{XieFP024GradSafe, QianZS025HSF, ChenXJLTG25jailbreakfree}. 
However, prior studies have shown that exposing internal signals (e.g., hidden states and gradients) could be inverted to recover inputs, training samples through reconstruction attacks~\cite{morris2023text,wan2024information,zhu2019deep}. 
These limitations raise a key research challenge: \emph{How to audit harmful-risk behaviors without inspecting raw user content or exposing model-internal semantic representations.}

To overcome this challenge, we exploit a previously \emph{underexplored auditing surface} exposed by MoE itself. 
MoE models exhibit input-dependent execution through sparse routing: input-dependent token representations are routed to selected experts, producing token-to-expert assignments and per-expert workloads.
When executed on GPUs, these workloads manifest as measurable low-level execution signals.
Throughout this paper, we use \emph{expert routing telemetry} to denote these routing-induced GPU execution signals rather than software-visible router logits or token-level routing assignments.
Building on this insight, we propose \textsc{RouteScan}, a non-intrusive auditing framework that detects harmful-risk inputs without accessing user content or model-internal states.

Specifically, \textsc{RouteScan} instantiates this execution-side auditing principle by monitoring the number of active GPU threads allocated to expert-module execution during the prefilling phase. 
The resulting thread-level telemetry serves as a compact hardware-side proxy for routing-induced expert workloads, without exposing prompts, outputs, hidden states, or gradients. 
To transform raw telemetry into robust auditing signals, \textsc{RouteScan} constructs two complementary feature views: fine-grained expert-load patterns that characterize workload distributions across expert modules, and coarse-grained layer-level statistics that capture the structural behavior of MoE execution. 
\textsc{RouteScan} then applies a Hybrid Feature Scoring and Adaptive Support mechanism to select invariant and discriminative feature subsets that remain stable across harmful domains and jailbreak wrappers. 
Finally, a lightweight linear detector with source-side score calibration
provides a consistent operating point within each deployment profile.

We conduct extensive experiments on four widely used MoE models with distinct routing designs
across heterogeneous hardware platforms.
Experimental results demonstrate that \textsc{RouteScan} achieves high detection accuracy and robust transferability across unseen harmful domains and novel jailbreak wrappers. 
Furthermore, privacy stress tests show that the collected telemetry does not support reliable exact prompt reconstruction or exact sensitive-field recovery under the evaluated attacks, although limited coarse-grained attribute information can remain measurable. 

The main contributions can be summarized as follows:

\begin{itemize}
    \item \textbf{Mechanism}: We, \emph{for the first time}, identify a previously underexplored execution-side auditing surface exposed by MoE models. 
    MoE models rely on a sparse routing mechanism that assigns input-dependent token representations to selected experts, inducing expert workload patterns that are reflected in low-level GPU execution telemetry. We show that these workload patterns could provide a non-content signal for auditing harmful-risk inputs without accessing user prompts or model-internal states.

    \item \textbf{Framework}: We propose \textsc{RouteScan}, a non-intrusive auditing framework that identifies harmful-risk inputs to MoE models from GPU-level execution telemetry, \emph{without accessing user content or model-internal states}. \textsc{RouteScan} uses active GPU-thread allocation during the prefilling phase as a proxy for routing-induced expert workloads, extracts hierarchical features from expert-load and layer-level execution patterns, and applies a Hybrid Feature Scoring and Adaptive Support mechanism with calibrated linear detection for robust auditing.
    
    \item \textbf{Evaluation}: We conduct extensive evaluations on open-source MoE models with distinct routing designs, covering four harmful-content benchmarks, three jailbreak wrappers, and heterogeneous GPU architectures. 
    \textsc{RouteScan} demonstrates strong transferability, achieving AUROC above $0.91$ on unseen harmful domains and above $0.95$ against new jailbreak wrappers. 
    Exact-content reconstruction tests further show that aggregated GPU telemetry collected during prefilling does not support reliable recovery of full prompts or exact sensitive fields under the evaluated attack, delineating an empirical boundary for content recovery.

\end{itemize}

\section{Background}
\label{sec:background}

\subsection{MoE Routing and Prefill Expert Loads}
\label{subsec:moe_arch}

Mixture-of-Experts (MoE) architectures scale large language models (LLMs) by replacing dense Feed-Forward Network (FFN) blocks with multiple expert networks while activating only a small subset of them for each token~\cite{FedusZS22,LepikhinLXCFHKS21GShard}. 
This sparse conditional-computation mechanism increases model capacity without proportionally increasing the per-token computational cost.

Let $\boldsymbol{h}_{l,t} \in \mathbb{R}^{d}$ denote the input hidden state of the $t$-th prompt token at the $l$-th MoE layer. 
A gating network produces the router logits
$\boldsymbol{z}_{l,t} \in \mathbb{R}^{M_l}$ over $M_l$ candidate experts and selects the index set
\begin{equation}
    \mathcal{S}_{l,t}
    =
    \operatorname{TopK}\!\left(\boldsymbol{z}_{l,t}, K_l\right),
\end{equation}
where $K_l \ll M_l$ is the number of experts activated for each token~\cite{ShazeerMMDLHD17}. 
For each selected expert $i \in \mathcal{S}_{l,t}$, the normalized routing weight and the resulting MoE output are
\begin{equation}
    g_{l,t,i}
    =
    \frac{\exp(z_{l,t,i})}
    {\sum_{j \in \mathcal{S}_{l,t}}\exp(z_{l,t,j})},
    \qquad
    \widetilde{\boldsymbol{h}}_{l,t}
    =
    \sum_{i \in \mathcal{S}_{l,t}}
    g_{l,t,i} E_i^{(l)}(\boldsymbol{h}_{l,t}),
\end{equation}
where $E_i^{(l)}$ denotes the $i$-th expert at layer $l$.

During the prefill phase, the model processes all $T$ prompt tokens before the decoding begins. 
The expert activation load at layer $l$ is defined as
\begin{equation}
    \ell_i^{(l)}
    =
    \sum_{t=1}^{T}
    \mathbb{I}\!\left[i \in \mathcal{S}_{l,t}\right],
    \qquad
    \boldsymbol{\ell}^{(l)}
    =
    \left[
    \ell_1^{(l)}, \ldots, \ell_{M_l}^{(l)}
    \right],
\end{equation}
where $\ell_i^{(l)}$ is the number of prompt tokens routed to expert $E_i^{(l)}$. 
In the absence of token dropping, the total number of expert assignments satisfies
$\sum_{i=1}^{M_l}\ell_i^{(l)} = T K_l$; therefore, the informative signal lies primarily in how these assignments are distributed across experts.

Given the input-dependent nature of expert routing, different prompts 
induce different expert-load distributions, which are reflected in 
low-level GPU execution behaviors such as active thread counts. 
\textsc{RouteScan} exploits these execution variations as non-content signals 
for harmful-risk auditing.

\subsection{Routing-Induced GPU Execution Telemetry}
\label{subsec:hardware_telemetry}

GPU execution telemetry consists of low-level runtime signals exposed by profiling and performance-monitoring interfaces, such as kernel events, thread activity, and memory transactions~\cite{YanLXQGCZ23MERCURY,Naghibijouybari18RenderedInsecure}. In \textsc{RouteScan}, we use only the subset of these signals that reflects routing-induced expert workloads; this hardware-derived subset is what we call \emph{expert routing telemetry}. 
Although prior work has exploited such signals as side channels to infer sensitive workload information~\cite{zhang2025nvbleedcovertsidechannelattacks,DingXSDF25MoEcho}, we use them for defensive safety auditing.
During the prefilling phase, input-dependent routing forms expert-specific token workloads, which are reflected in the thread activity of expert-execution kernels. 
Prior work has shown that GPU thread counts are closely correlated with expert activation loads~\cite{DingXSDF25MoEcho}. 
Accordingly, \textsc{RouteScan} collects per-expert thread counts as hardware-derived expert routing telemetry to characterize routing-induced expert workloads for harmful behavior auditing.

\section{Harmful Behavior Auditing Paradigm}
\subsection{Input Taxonomy}
\label{subsec:input_taxonomy}
We formalize harmful behavior auditing by categorizing the prompt space into three classes based on user intent:

\noindent\textbf{Benign Prompts}
represent safe and legitimate user queries that aim to obtain ordinary information or services in full compliance with service providers' policies.

\noindent\textbf{Harmful Prompts}
refer to inputs that explicitly and directly request content that violates policies or is malicious, whose intents typically include generating hate speech, disinformation, malicious software code, and so forth \cite{weidinger2022taxonomy, wang2023donotanswer}. 

\noindent\textbf{Jailbreak Prompts} 
refer to adversarially crafted inputs designed to subvert LLM safety mechanisms. By employing techniques such as role-playing and complex instruction obfuscation~\cite{Liu2023AutoDANGS, chao2025jailbreakingPAIR, Lv2024CodeChameleonPE}, attackers disguise malicious intent to bypass safety guardrails and induce the generation of harmful content.

\subsection{Threat Model}
\label{subsec:threat_model}

We consider a practical deployment scenario: a cloud-hosted MoE model service managed by a cloud service provider. 
The provider may need to audit harmful-risk behavior for internal compliance, platform governance, or external regulatory review. 
However, directly exposing user prompts, model outputs, or model internals to the auditing process can raise serious concerns about user privacy and model confidentiality. 
To capture this privacy-constrained auditing setting, we define a tripartite ecosystem consisting of \textbf{Users}, \textbf{Service Providers}, and \textbf{Regulators}. 
The auditing interface follows a least-privilege design, exposing only 
aggregated routing-induced GPU execution telemetry collected during the prefilling phase, while hiding 
raw prompts, model outputs, and internal model states.

\noindent\textbf{Users: }
Users interact with the MoE model via application layer APIs. Most users submit benign queries, whereas malicious users may craft harmful or jailbreak prompts to bypass safety alignments. 
We model malicious users as \textit{prompt-level adversaries}: they can control input prompts and observe model responses, but they have no access to model parameters, service infrastructure, or hardware telemetry.

\noindent\textbf{Service Providers: }
Providers manage the MoE inference infrastructure and enforce existing safety mechanisms to handle user requests online, such as allowing or rejecting them. Meanwhile, they collect request-level aggregated GPU execution telemetry during the prefilling phase as input to external auditing. 
We model service providers as \emph{audit-cooperative but disclosure-constrained}. They follow the prescribed auditing protocol and provide external auditors with the information necessary to conduct the audit. However, to prevent privacy-sensitive information, such as plaintext prompts and model outputs, from being further disclosed beyond the serving plane, they release only aggregated GPU telemetry and necessary audit metadata through a predefined auditing interface.

\begin{figure}[t]
  \centering
  \includegraphics[width=\columnwidth]
  {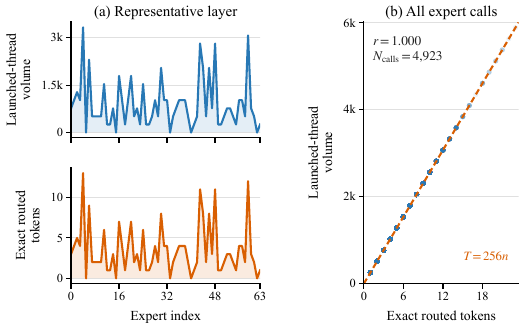}
  \caption{Validation of the execution measurement chain on OLMoE/RTX~5880.
  \textbf{(a)} For a deterministically selected representative layer,
  software-recorded routed-token counts and Nsight-derived launched-thread
  volumes exhibit the same expert-wise pattern.
  \textbf{(b)} Each expert kernel call is plotted by its routed-token count
  and launched-thread volume. All points lie along a linear reference line,
  indicating a linear relationship between the two measurements. The 4,923
  calls occupy only 23 distinct workload coordinates; darker markers indicate
  more calls at the corresponding coordinate. Inactive experts are omitted.}
  \label{fig:workload_telemetry_alignment}
\end{figure}

\begin{figure}[t]
  \centering
  \includegraphics[width=\columnwidth]{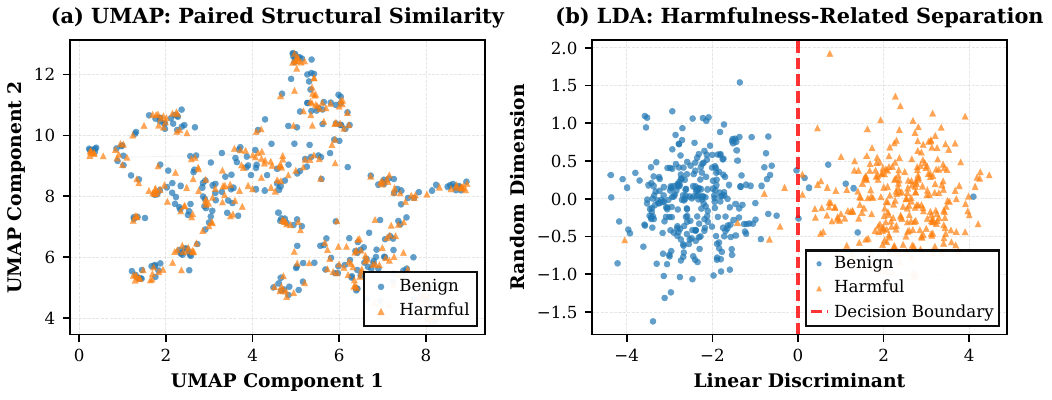}
  \caption{Comparison of unsupervised and supervised dimensionality reduction on paired benign--harmful samples. \textbf{(a)} UMAP projects paired benign--harmful samples into nearby regions, consistent with the controlled construction in which each pair shares the same syntactic template and differs mainly in risk-related keywords. \textbf{(b)} LDA, however, recovers a clear separating direction, showing that MoE activation features encode harmfulness-related signals beyond surface syntactic structure.}
  \label{fig:umap_lda}
\end{figure}

\noindent\textbf{Regulators: }
Regulators aim to enforce safety guidelines through a post-hoc accountability mechanism under strict privacy constraints. 
They are not granted access to raw user prompts, model outputs, or query-level semantic information. 
Instead, their observations are limited to aggregated GPU execution telemetry reported by service providers. 
In addition, we assume regulators have access to labeled traces collected from controlled auditing workloads for profiling telemetry patterns of benign and harmful requests.

\begin{figure*}[t]
    \centering
    \includegraphics[width=0.95\linewidth]{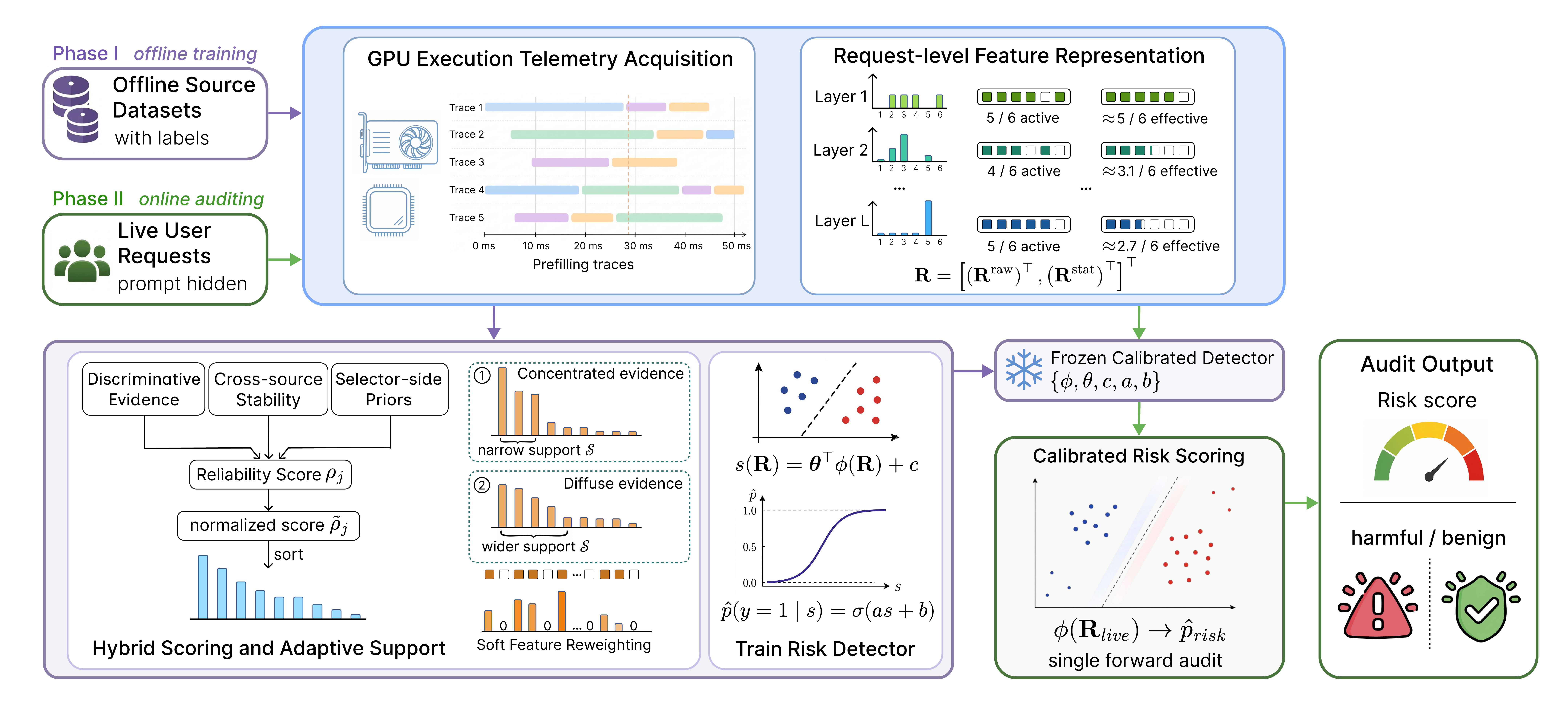}
    \caption{Overview of \textsc{RouteScan}. \textsc{RouteScan} performs harmful-behavior auditing for MoE LLM services by collecting aggregated GPU execution telemetry associated with expert execution during the prefilling phase. Without accessing raw user prompt text, model responses, or internal neural representations, it constructs a request-level telemetry representation, uses hybrid feature scoring and adaptive support selection to emphasize robust telemetry dimensions, and outputs a source-calibrated harmful-risk score through a lightweight linear detector.}    
    \label{fig:overview}
\end{figure*}

\section{Motivation and Preliminary Evidence}
\label{sec:motivation}

\subsection{Motivation for MoE Execution-Side Auditing}

MoE models exhibit an input-dependent sparse routing mechanism, where 
different prompts induce different expert activation patterns and workload 
distributions. These routing behaviors are reflected in observable GPU 
execution characteristics, such as expert-level thread activity. Prior 
studies have further shown that input-dependent expert execution leaves 
measurable traces on CPU and GPU platforms~\cite{DingXSDF25MoEcho}. These 
observations motivate us to investigate whether routing-induced execution 
telemetry can serve as an effective signal for harmful behavior auditing 
without directly accessing raw prompts, model outputs, or internal model 
representations.

\subsection{Preliminary Telemetry Analysis}

We validate the execution measurement chain on OLMoE deployed on an RTX~5880.
During prefilling, the software wrapper records routed-token counts per expert,
while Nsight captures corresponding expert-kernel launch dimensions. We align
the measurements via CUDA correlation IDs and derive launched-thread volume
from kernel grid/block dimensions. Figure~\ref{fig:workload_telemetry_alignment}
shows a strong linear correlation between launched-thread volume and routed
workload, confirming GPU telemetry as a proxy for expert execution load.

We further evaluate telemetry separability on HEx-PHI~\cite{anonymous2024finetuninghexphi}.
We construct paired benign prompts by replacing harmful expressions while
preserving prompt structures. Figure~\ref{fig:umap_lda} compares telemetry
distributions using UMAP and LDA. UMAP shows substantial benign-harmful overlap,
while LDA achieves \(F_1=0.892\), indicating that execution features contain
harmfulness-related signals. These results support the feasibility of
GPU telemetry-based harmful behavior auditing.

\section{Method}
\label{sec:method}

\subsection{System Overview}
\label{sec:arch}

As illustrated in Figure~\ref{fig:overview}, \textsc{RouteScan} is a request-level safety auditing system for MoE LLM services. Rather than accessing plaintext user prompts, model responses, or internal neural representations, \textsc{RouteScan} estimates the safety risk of each request from aggregated routing-induced GPU execution telemetry collected during the prefilling phase.

In the first stage, \textsc{RouteScan} extracts hardware-derived expert routing telemetry during the prefilling phase of each LLM request and converts it into a request-level telemetry representation. Fine-grained expert-load patterns are combined with coarse-grained layer-level structural statistics, yielding a unified GPU-execution-layer characterization of each request.

The second stage introduces a hybrid feature scoring and adaptive support 
mechanism. Each telemetry dimension is scored by jointly considering 
discriminative evidence, cross-source stability, and conservative selector-side priors; based on the concentration of feature scores, the support selector adaptively determines how many dimensions to retain and applies soft reweighting to the selected dimensions, improving transferability to unseen harmful domains and robustness to novel jailbreak wrappers.

In the final stage, a lightweight linear detector is trained on the selected robust features. Source-side adaptive regularization accounts for differences in margin structure across deployment profiles, while score calibration maps raw detector margins onto a calibrated harmful-risk scale. The output is a source-calibrated harmful-risk score for each request under the given deployment profile.

\subsection{Request-level Feature Representation}
\label{sec:telemetry_repr}

To provide a unified and discriminative physical-layer telemetry input for subsequent harmful-risk auditing, \textsc{RouteScan} constructs a feature representation $\mathbf{R}$ for each inference request, designed to capture both fine-grained expert-level load details and coarse-grained layer-level structural statistics.

\paragraph{Expert Load Distribution.}
At the fine-grained level, the system focuses on the workload observed for each expert. Let $n_{l,e}$ denote the aggregated execution load recorded by GPU telemetry for expert $e$ in layer $l$ during the prefilling phase of the current request. We define the normalized load distribution of layer $l$ as
\begin{equation}
p_{l,e}=\frac{n_{l,e}}{\sum_{e'=1}^{E_l} n_{l,e'}}, 
\qquad
\mathbf{p}_l=\left[p_{l,1},\ldots,p_{l,E_l}\right]^{\top}.
\label{eq:raw-routing}
\end{equation}
Normalizing by the total layer load preserves the relative load structure within each layer while mitigating the impact of global-scale variations including request length and overall load magnitude.

\paragraph{Layer-level Structural Statistics.}
Fine-grained expert-load distributions retain detailed routing information for individual experts, but do not explicitly summarize the overall execution structure within each MoE layer. To complement these expert-level features, we derive coarse-grained layer-level statistics that characterize routing coverage, effective spread, and load concentration.

First, we define the \textit{active expert count} for layer $l$ as
\begin{equation}
a_l = |\{e : n_{l,e}>0\}|,
\label{eq:active-count}
\end{equation}
and normalize it into the \textit{activation coverage rate}
\begin{equation}
v_{\mathrm{act}}^{(l)}=\frac{a_l}{E_l}.
\label{eq:active-rate}
\end{equation}
The activation coverage rate \(v_{\mathrm{act}}^{(l)}\) measures the fraction of experts touched by the request in layer \(l\), capturing the breadth of expert execution.

Second, we define the \textit{effective expert count} for layer $l$ as
\begin{equation}
\tilde{a}_l = \exp(H(\mathbf{p}_l)),
\label{eq:eff-count}
\end{equation}
where $H(\mathbf{p}_l)$ denotes the Shannon entropy of the normalized load distribution in layer $l$, computed with the natural logarithm and the convention that $0\log 0=0$. Based on this, we further define the \textit{effective activation rate}
\begin{equation}
v_{\mathrm{eff}}^{(l)}=\frac{\tilde{a}_l}{E_l}.
\label{eq:eff-rate}
\end{equation}
Unlike $v_{\mathrm{act}}^{(l)}$, which treats all touched experts equally regardless of their load, $v_{\mathrm{eff}}^{(l)}$ reflects the effective spread of expert execution: a request that heavily concentrates load on two experts yields a much lower effective count than one that distributes load evenly across many experts.

Building upon the aforementioned statistics, we introduce two derived structural residuals. The \textit{coverage gap} for layer \(l\) is defined as
\begin{equation}
g_{\mathrm{cov}}^{(l)} = v_{\mathrm{act}}^{(l)} - v_{\mathrm{eff}}^{(l)} ,
\label{eq:coverage-gap}
\end{equation}
which measures the difference between the routing breadth and the entropy-based effective spread. 

The \textit{coverage concentration} characterizes the relative contraction from the set of touched experts to the effective load width, and is computed as
\begin{equation}
c_{\mathrm{cov}}^{(l)} = 1 - \frac{v_{\mathrm{eff}}^{(l)}}
{v_{\mathrm{act}}^{(l)}+\epsilon_{\mathrm{cov}}},
\label{eq:coverage-concentration}
\end{equation}
where \(\epsilon_{\mathrm{cov}}>0\) is a small constant for numerical stability. A larger value indicates that the load among touched experts is concentrated on a narrower effective subset.

\paragraph{Unified Representation.}
Finally, we concatenate the fine-grained expert-load distributions and the
coarse-grained layer-level structural statistics to form a unified request-level feature representation. For each layer \(l\), we define the structural statistic vector as
\(\mathbf{s}^{(l)} =
[v_{\mathrm{act}}^{(l)}, v_{\mathrm{eff}}^{(l)},
g_{\mathrm{cov}}^{(l)}, c_{\mathrm{cov}}^{(l)}]^{\top}\).
The raw expert-load block and the structural-statistics block are then constructed by concatenating the corresponding layer-wise vectors in increasing layer order:
\begin{equation}
\begin{aligned}
    \mathbf{R}^{\mathrm{raw}}
    &=
    \left[
        \mathbf{p}_1^{\top},
        \ldots,
        \mathbf{p}_L^{\top}
    \right]^{\top}, \\
    \mathbf{R}^{\mathrm{stat}}
    &=
    \left[
        \left(\mathbf{s}^{(1)}\right)^{\top},
        \ldots,
        \left(\mathbf{s}^{(L)}\right)^{\top}
    \right]^{\top}.
\end{aligned}
\label{eq:r-blocks}
\end{equation}
The final representation concatenates the two blocks:
\begin{equation}
    \mathbf{R}
    =
    \left[
        \left(\mathbf{R}^{\mathrm{raw}}\right)^{\top},
        \left(\mathbf{R}^{\mathrm{stat}}\right)^{\top}
    \right]^{\top}.
    \label{eq:unified-feature}
\end{equation}
The resulting vector has dimension \(d = \sum_{l=1}^{L}E_l+4L\) and is used as the
input to the subsequent hybrid scoring and adaptive support module.

\subsection{Hybrid Scoring and Adaptive Support}
\label{sec:selector_family}

Although the unified representation combines raw load distributions with
structural telemetry summaries, its dimensions vary in cross-source reliability. Treating all dimensions equally can amplify source-specific variation, while fixed-width truncation can remove transferable tail dimensions.
To address both failure modes, \textsc{RouteScan} adopts a hybrid scoring and adaptive support framework: the system first estimates the reliability of each telemetry dimension as cross-source harmful-risk evidence, then adaptively determines the width of the support set according to the score distribution, and finally applies a target-independent soft reweighting within the selected support.

\paragraph{Dimension-level Hybrid Scoring.}
For each feature dimension \(j\), \textsc{RouteScan} assigns a reliability score \(\rho_j\) by combining discriminative evidence, cross-source stability, and conservative selector-side priors:
\begin{equation}
\rho_j
=
\mathcal{M}_{\mathrm{disc}}(j)\,
\mathcal{M}_{\mathrm{cons}}(j)\,
\mathcal{M}_{\mathrm{prior}}(j).
\label{eq:hybrid-score}
\end{equation}
Here, \(\mathcal{M}_{\mathrm{disc}}(j)\) captures the discriminative reliability of dimension \(j\), combining pooled single-dimension separability with cross-source invariance, so that a dimension receives a high score only when it separates harmful and benign requests while preserving a stable harmful--benign gap across source domains.
\(\mathcal{M}_{\mathrm{cons}}(j)\) measures whether the feature exhibits stable directional behavior across source domains and resampling trials, thereby suppressing dimensions that frequently flip sign or are overly sensitive to sampling fluctuations.
\(\mathcal{M}_{\mathrm{prior}}(j)\) is a selector-side adjustment factor that encodes structural preferences and boundary-sensitivity controls. It includes a mild layer-dependent prior to avoid over-concentrating the support on potentially source-specific late-layer features, and a boundary-sensitivity penalty that downweights dimensions whose discriminative signal is driven primarily by edge-positive samples rather than prototypical harmful inputs.

Subsequently, the system applies non-negative truncation and normalization to obtain the normalized score $\tilde{\rho}_j$ and the corresponding score-mass value $\pi_j$:
\begin{equation}
\tilde{\rho}_j=
\frac{\max(\rho_j,0)}
{\max_k \max(\rho_k,0)},
\qquad
\pi_j=
\frac{\tilde{\rho}_j}
{\sum_{k:\tilde{\rho}_k>0}\tilde{\rho}_k}.
\label{eq:hybrid-score-norm-mass}
\end{equation}

\paragraph{Entropy-based Adaptive Support.}
Let \(\boldsymbol{\pi}=(\pi_1,\ldots,\pi_d)^{\top}\) denote the score-mass
vector. If high scores are concentrated in only a few dimensions,
\(\boldsymbol{\pi}\) becomes sharper; if many dimensions carry signals of
moderate strength, it becomes more diffuse. Analogous to the construction of the effective expert count in Section~\ref{sec:telemetry_repr}, we use the Shannon entropy of the score-mass distribution to quantify the dispersion of the effective harmful-risk signal:
\begin{equation}
\delta=
\frac{\exp(H(\boldsymbol{\pi}))}
{|\{j:\tilde{\rho}_j>0\}|}.
\label{eq:score-diffuseness}
\end{equation}

The system then transforms this dispersion into a target cumulative mass $q(\delta)$ through a sigmoid mapping. By sorting all dimensions in descending order of $\tilde{\rho}_j$, the system selects the smallest prefix whose cumulative mass is no less than $q(\delta)$, and defines it as the support set $\mathcal{S}$. Consequently, when the harmful-risk signal is more diffuse, the system automatically retains a broader support set; when the signal is more concentrated, the support set naturally shrinks.

\paragraph{Soft Feature Reweighting.}
Let \(\mathbf{w}=(w_1,\ldots,w_d)^{\top}\) denote the soft weights assigned to the \(d\) original telemetry dimensions. For each dimension retained in the selected support, the system derives a lower-bounded weight from its normalized feature score, while assigning zero weight to all dimensions outside the support:
\begin{equation}
w_j=
\begin{cases}
\eta + (1-\eta)\tilde{\rho}_j^{\,\kappa}, & j\in\mathcal{S},\\
0, & j\notin\mathcal{S}.
\end{cases}
\label{eq:mass-soft-operator}
\end{equation}
where \(\eta\) controls the lower bound within the selected support and
\(\kappa\) controls the curvature of score-based soft reweighting.
The weights \(w_j\) control the influence of selected telemetry dimensions in the linear detector, while dimensions outside \(\mathcal{S}\) are excluded.

\subsection{Linear Risk Detection and Score Calibration}
\label{sec:detector_calibration}
\paragraph{Detector-input Transformation.}
For linear detection, \textsc{RouteScan} constructs the detector input
\(\phi(\mathbf{R})\) from the representation \(\mathbf{R}\) through a
source-fitted feature transformation.
The transformation \(\phi(\cdot)\) first restricts the feature space to the selected telemetry dimensions in \(\mathcal{S}\), so that non-selected
dimensions do not enter the detector input. It then applies column-wise max-absolute scaling fitted on the source training split, followed by column reweighting using the selector soft weights and fixed block-wise weights. All components of this transformation are determined from source data and remain fixed during target evaluation.
The block-wise weights control the relative influence of different feature
blocks in the linear detector. Fine-grained expert-load dimensions remain the primary discriminative evidence, while layer-level basic statistics and structural residuals serve as auxiliary signals. This design allows
structural execution-shape information to complement raw expert-load patterns without allowing low-dimensional structural summaries to dominate the linear boundary. 

\paragraph{Linear Detector.}
Building on the transformed detector input, we employ a lightweight linear detector for harmful-risk discrimination. The raw detector, which corresponds to the uncalibrated linear output before score calibration, is
\begin{equation}
s(\mathbf{R}) =
\boldsymbol{\theta}^{\top}\phi(\mathbf{R}) + c,
\label{eq:linear-score}
\end{equation}
where \(\boldsymbol{\theta}\) and \(c\) are learned by logistic regression on the transformed source training data.

We adopt a linear detector by design, aiming to concentrate the primary modeling capacity on telemetry representation construction and robust feature support selection, rather than on a high-capacity classifier that may overfit source-specific, higher-order patterns. 

\paragraph{Source-adaptive Regularization.}
Different deployment profiles and risk-positive compositions can alter the scale and distribution of source-side detector margins. 
A fixed logistic regularization strength applied uniformly across all settings may fail to accommodate source configurations with varying degrees of separability. 
To address this issue, \textsc{RouteScan} adaptively selects the inverse \(\ell_2\)-regularization strength \(C\) using only margin statistics from the source validation split. This source-side adaptation balances the preservation of harmful-risk discriminative information against the suppression of source-specific patterns.

To compute these statistics, the system first fits a temporary linear detector on the source training data using a reference strength \(C_{\mathrm{ref}}\), and computes raw margins on the source validation set. Since the source positive pool may contain heterogeneous risk-positive samples, relying only on the pooled positive mean margin can obscure weakly separated subsets. The adaptive regularizer therefore evaluates source-side margins over protocol-defined positive subsets.

The risk-positive validation samples are partitioned into a protocol-defined collection of disjoint positive subsets \(\mathcal{P}\). Let \(\mu_0\) denote the average margin of the benign validation samples, and let \(\mu_P\) denote the average margin of a positive subset \(P \in \mathcal{P}\). We define the weakest positive separation \(\Delta_{\mathrm{sep}}\) and the positive-subset disparity \(r_{\mathrm{sub}}\) as follows:
\begin{equation}
\begin{aligned}
  \Delta_{\mathrm{sep}}
  &=
  \min_{P \in \mathcal{P}}(\mu_P - \mu_0), \\
  r_{\mathrm{sub}}
  &=
  \frac{
     \max_{P \in \mathcal{P}} \mu_P - \min_{P \in \mathcal{P}} \mu_P
   }{|\Delta_{\mathrm{sep}}| + \epsilon_{\mathrm{sep}}}.
\end{aligned}
\label{eq:adaptive-c-source-stats}
\end{equation}
Here, \(\Delta_{\mathrm{sep}}\) measures the margin distance between the benign group and the hardest-to-separate positive subset, while \(r_{\mathrm{sub}}\) captures the margin disparity among positive subsets, normalized by the scale of the weakest separation. For example, in the mixed-positive protocol, direct harmful prompts and AutoDAN jailbreak prompts form separate positive subsets; in the harmful-only protocol, \(\mathcal{P}\) contains a single positive subset and \(r_{\mathrm{sub}}=0\).

The system then selects the final regularization parameter through a bounded monotonic mapping:
\begin{equation}
  C = G(\Delta_{\mathrm{sep}},\, r_{\mathrm{sub}}).
  \label{eq:adaptive-c-main}
\end{equation}
The mapping \(G\) increases with stronger positive-negative separation and
decreases with larger positive-subset disparity. When the source validation margins show strong separation and the positive subsets are consistently separated from benign samples, the system selects a larger \(C\), allowing the linear boundary to retain more discriminative detail. Conversely, when the weakest separation is small or the subset disparity is large, the system selects a smaller \(C\), imposing stronger regularization to suppress source-specific overfitting.

\paragraph{Score Calibration.}
Finally, although the raw detector margin \(s(\mathbf{R})\) provides a useful ranking score, its absolute scale depends on the source training distribution, feature scaling, and the selected regularization strength. As a result, the same margin value may not correspond to the same operating point across deployment profiles. \textsc{RouteScan} therefore calibrates the raw margin on the source validation set. We use a one-dimensional Platt mapping~\cite{platt1999probabilistic}:
\begin{equation}
\hat{p}(y=1 \mid s) = \sigma(a s + b),
\label{eq:platt}
\end{equation}
where the parameters \((a,b)\) are estimated exclusively from the raw margins and binary labels on the source validation set. During deployment, each request is first scored by the linear detector to obtain \(s(\mathbf{R})\), and is then mapped to a source-calibrated harmful-risk score \(\hat{p}\).

\section{Evaluation}
\label{sec:eval}

\subsection{Evaluation Protocols and Setup}
\label{sec:eval_setup}

\paragraph{Models and hardware.}
We evaluate \textsc{RouteScan} under multiple model/hardware deployment profiles to examine whether the telemetry pipeline remains effective beyond a single configuration. 
In our experiments, these profiles are instantiated with four open-source MoE LLMs, DeepSeek-V2-Lite-Chat \cite{liu2024deepseekV2}, Qwen1.5-MoE-A2.7B-Chat \cite{yang2024qwen2}, GPT-OSS-20b\cite{openai2025gptoss}, OLMoE-1B-7B-0125-Instruct\cite{muennighoff2025olmoe}. We deploy these models on two different types of GPUs: NVIDIA RTX A6000 and NVIDIA RTX 5880.

\paragraph{Datasets.}
Following the taxonomy in Section~\ref{subsec:input_taxonomy}, we distinguish benign prompts, harmful prompts, and jailbreak prompts. 
Harmful prompts are sourced from multiple harmful-content benchmarks, including AdvBench~\cite{zou2023universaladvbench}, StrongREJECT~\cite{SoulyLBTHPASEWT24strongreject}, HEx-PHI~\cite{anonymous2024finetuninghexphi}, CatHarmfulQA~\cite{bhardwaj2024languagecatharmful}, and HarmBench-Standard~\cite{mazeika2024harmbench}. 
Jailbreak prompts are generated by applying jailbreak wrappers to harmful prompts and are used in the mixed-positive family transfer protocol.

For the benign negative class, we construct paired benign prompts by
replacing malicious keywords or entities in harmful prompts with safe
alternatives while preserving the original phrasing.
This process ensures high structural similarity between paired samples, effectively isolating the harmful intent as the sole distinguishing variable.
We enforce strict prompt-group-level isolation between training and testing splits. Specifically, all variants of a given prompt are restricted exclusively to either the training or the testing set, ensuring no data leakage.

\paragraph{Cross-domain harmful auditing.}
To evaluate the generalization of prefilling-phase GPU telemetry for auditing harmful behaviors across unseen domains, we employ a leave-one-target-out protocol. 
In each fold, one dataset is held out as the unseen target domain, and the remaining datasets constitute the source pool. The entire training pipeline is performed on the source pool, leaving the target data for evaluation only.
We define harmful prompts as the positive class and their paired benign counterparts as the negative class.

\paragraph{Mixed-positive family transfer.}
The mixed-positive family transfer protocol evaluates generalization to unseen jailbreak wrappers. 
Under the auditing objective, both harmful prompts and jailbreak prompts are treated as risk-positive requests, because both correspond to harmful-risk behavior, while they differ in surface realization. 
Specifically, the source risk-positive pool contains original harmful prompts and AutoDAN~\cite{Liu2023AutoDANGS} jailbreak prompts that successfully elicit unsafe model outputs from the same benchmark, with paired benign prompts serving as the negative class. 
During testing, the risk-positive class is replaced by jailbreak prompts generated with an unseen attack wrapper, and the negative class consists of the corresponding held-out paired benign prompts from the same benchmark. 
The protocol thereby investigates whether \textsc{RouteScan} can still identify harmful-risk behavior from GPU execution telemetry when the same harmful intent is repackaged by a novel jailbreak strategy.

\begin{table*}[t]
\centering
\caption{
Cross-domain harmful auditing results.
}
\label{tab:harmful-main}
\resizebox{\textwidth}{!}{
\begin{tabular}{llccc ccc ccc ccc}
\toprule
\multirow{2}{*}{Model} & \multirow{2}{*}{Hardware}
& \multicolumn{3}{c}{AdvBench}
& \multicolumn{3}{c}{StrongReject}
& \multicolumn{3}{c}{CatHarmfulQA}
& \multicolumn{3}{c}{HarmBench} \\
\cmidrule(lr){3-5}
\cmidrule(lr){6-8}
\cmidrule(lr){9-11}
\cmidrule(lr){12-14}
& & AUROC & F1@0.5 & Acc@0.5
& AUROC & F1@0.5 & Acc@0.5
& AUROC & F1@0.5 & Acc@0.5
& AUROC & F1@0.5 & Acc@0.5 \\
\midrule

\multirow{2}{*}{DeepSeek} & A6000
& 0.9984 & 0.9528 & 0.9548
& 0.9760 & 0.8908 & 0.8978
& 0.9798 & 0.9086 & 0.9127
& 0.9865 & 0.8989 & 0.8875 \\
& 5880
& 0.9985 & 0.9561 & 0.9577
& 0.9833 & 0.9057 & 0.9105
& 0.9817 & 0.9219 & 0.9246
& 0.9927 & 0.9153 & 0.9075 \\
\midrule

\multirow{2}{*}{Qwen} & A6000
& 0.9888 & 0.9440 & 0.9462
& 0.9713 & 0.9167 & 0.9137
& 0.9512 & 0.8848 & 0.8818
& 0.9816 & 0.9231 & 0.9175 \\
& 5880
& 0.9639 & 0.9207 & 0.9250
& 0.9803 & 0.9221 & 0.9201
& 0.9504 & 0.8823 & 0.8809
& 0.9814 & 0.9151 & 0.9100 \\
\midrule

\multirow{2}{*}{OLMoE} & A6000
& 0.9910 & 0.9376 & 0.9404
& 0.9670 & 0.8988 & 0.9026
& 0.9189 & 0.8471 & 0.8455
& 0.9846 & 0.9234 & 0.9200 \\
& 5880
& 0.9985 & 0.9845 & 0.9846
& 0.9841 & 0.9316 & 0.9345
& 0.9336 & 0.8713 & 0.8700
& 0.9911 & 0.9469 & 0.9450 \\
\midrule

\multirow{2}{*}{GPT-OSS} & A6000
& 0.9959 & 0.9261 & 0.9308
& 0.9828 & 0.9272 & 0.9297
& 0.9276 & 0.8616 & 0.8536
& 0.9877 & 0.9163 & 0.9100 \\
& 5880
& 0.9981 & 0.9489 & 0.9510
& 0.9861 & 0.9544 & 0.9553
& 0.9273 & 0.8663 & 0.8564
& 0.9876 & 0.9209 & 0.9150 \\

\bottomrule
\end{tabular}
}
\end{table*}

\paragraph{Privacy boundary evaluation.}
A central privacy concern is whether the signals used for harmful-risk detection could similarly facilitate input inversion or sensitive-attribute inference. To investigate this boundary, we design a medical privacy stress test utilizing synthetic clinical notes from the Asclepius study~\cite{KweonKKICBOLMYB24AsclepiusSynClinical} and the Synthetic Clinical Notes Embedded dataset~\cite{technoculture_synthetic_clinical_notes_embedded}. Although these sources contain no real patient records, they provide clinically structured synthetic text suitable for constructing privacy-sensitive scenarios. We employ a GPT-based pipeline to transform these notes into user-facing medical prompts, preserving coarse clinical attributes (e.g., disease category, severity) while defining explicit sensitive fields for recovery auditing.

In this setting, an attacker with access to the same aggregated prefilling-phase telemetry attempts to compromise user privacy via two complementary pathways: (1) \textit{telemetry-to-text inversion}, which seeks to reconstruct the original prompt and recover sensitive fields; and (2) \textit{telemetry-to-attribute probing}, which bypasses reconstruction to directly infer coarse-grained attributes (e.g., high-severity status) using binary probes trained on telemetry vectors.

\begin{table*}[t]
\centering
\caption{
Mixed-positive family transfer results.
}
\label{tab:jailbreak-main}
\renewcommand{\arraystretch}{1.05}
\resizebox{\textwidth}{!}{
\begin{tabular}{lllccc ccc ccc}
\toprule
\multirow{2}{*}{Model} & \multirow{2}{*}{Hardware} & \multirow{2}{*}{Source Benchmark}
& \multicolumn{3}{c}{AutoDAN}
& \multicolumn{3}{c}{PAIR}
& \multicolumn{3}{c}{CodeChameleon} \\
\cmidrule(lr){4-6}
\cmidrule(lr){7-9}
\cmidrule(lr){10-12}
& & & AUROC & F1@0.5 & Acc@0.5
& AUROC & F1@0.5 & Acc@0.5
& AUROC & F1@0.5 & Acc@0.5 \\
\midrule
\multirow{4}{*}{Qwen} & \multirow{2}{*}{A6000} & AdvBench + AutoDAN
& 1.0000 & 0.9952 & 0.9951
& 1.0000 & 0.9947 & 0.9946
& 0.9981 & 0.9442 & 0.9466 \\
& & StrongReject + AutoDAN
& 1.0000 & 0.9764 & 0.9758
& 0.9965 & 0.9735 & 0.9732
& 0.9987 & 0.9764 & 0.9758 \\
\cmidrule(lr){2-12}
& \multirow{2}{*}{5880} & AdvBench + AutoDAN
& 0.9999 & 0.9671 & 0.9660
& 0.9764 & 0.9662 & 0.9650
& 0.9712 & 0.9671 & 0.9660 \\
& & StrongReject + AutoDAN
& 1.0000 & 0.9612 & 0.9597
& 0.9808 & 0.9333 & 0.9322
& 1.0000 & 0.9612 & 0.9597 \\
\midrule
\multirow{4}{*}{DeepSeek} & \multirow{2}{*}{A6000} & AdvBench + AutoDAN
& 1.0000 & 0.9858 & 0.9856
& 0.9929 & 0.9712 & 0.9712
& 0.9848 & 0.9662 & 0.9663 \\
& & StrongReject + AutoDAN
& 1.0000 & 0.9265 & 0.9206
& 0.9540 & 0.8480 & 0.8468
& 0.9997 & 0.9265 & 0.9206 \\
\cmidrule(lr){2-12}
& \multirow{2}{*}{5880} & AdvBench + AutoDAN
& 1.0000 & 0.9858 & 0.9856
& 0.9949 & 0.9858 & 0.9856
& 0.9944 & 0.9858 & 0.9856 \\
& & StrongReject + AutoDAN
& 1.0000 & 0.9403 & 0.9365
& 0.9651 & 0.8976 & 0.8952
& 0.9942 & 0.9403 & 0.9365 \\
\bottomrule
\end{tabular}
}
\end{table*}

\subsection{Metrics and Calibration}
\label{sec:metrics_protocol}

Both detection protocols follow the same evaluation procedure. As described in Section~\ref{sec:detector_calibration}, raw detector margins are transformed into source-calibrated harmful-risk scores using a Platt mapping fitted on the source validation set. Fixed-threshold metrics are computed in this calibrated score space using a default threshold of \(0.5\). Target data are used only for final evaluation and never for training, feature selection, threshold selection, or calibration.

To quantify detection performance under both protocols, the main text reports three metrics. 
\(\mathrm{AUROC}\) measures ranking separability independent of any fixed threshold; \(\mathrm{F1}@0.5\) and \(\mathrm{Acc}@0.5\) evaluate performance at the fixed deployment threshold of 0.5: the former summarizes the precision-recall trade-off, and the latter reflects overall classification accuracy. The privacy boundary experiment addresses a fundamentally different question and is evaluated using dedicated inversion metrics, reported separately in Section~\ref{sec:eval_privacy}.

\subsection{Cross-domain Harmful Auditing}
\label{sec:eval_harmful}

Under the leave-one-target-out protocol defined in Section~\ref{sec:eval_setup}, Table~\ref{tab:harmful-main} reports the cross-domain harmful-auditing results. \textsc{RouteScan} achieves strong ranking separability across all evaluated model and hardware profiles, with all AUROC scores above \(0.91\). AdvBench and HarmBench exhibit particularly strong performance, while StrongReject remains stable across deployment profiles.

CatHarmfulQA is the most challenging target domain, particularly for OLMoE and GPT-OSS. The largest degradation occurs on OLMoE/A6000, where the AUROC decreases to \(0.9189\), suggesting a stronger target-domain shift in the induced telemetry patterns. Nevertheless, the fixed-threshold results remain competitive overall, indicating that source-calibrated telemetry features preserve transferable harmful-risk signals across unseen domains, despite residual domain- and deployment-specific shifts.

\subsection{Mixed-Positive Family Transfer}
\label{sec:eval_hjtransfer}

\begin{table*}[t]
\centering
\caption{
Telemetry-to-text inversion results on held-out medical privacy prompts.
}
\label{tab:privacy-inversion}
\resizebox{\textwidth}{!}{
\begin{tabular}{ccc ccc cccc c}
\toprule
\multicolumn{3}{c}{Prompt Reconstruction}
& \multicolumn{3}{c}{Exact Sensitive-Field Recovery}
& \multicolumn{4}{c}{Coarse Attributes from Generated Text}
& \multirow{2}{*}{LLM Sensitive-Fact Hit} \\
\cmidrule(lr){1-3}
\cmidrule(lr){4-6}
\cmidrule(lr){7-10}
Token F1 & Jaccard & ROUGE-L
& All Fields F1 & Identifier F1 & Condition F1
& Scenario F1 & Chronicity F1 & Severity F1 & Family F1
& \\
\midrule
0.1601 & 0.1116 & 0.1194
& 0.0000 & 0.0000 & 0.0000
& 0.2450 & 0.2110 & 0.2400 & 0.0420
& 0/54 \\
\bottomrule
\end{tabular}
}
\end{table*}
\begin{table*}[t]
\centering
\caption{
Targeted coarse-attribute probing from aggregated GPU execution telemetry.
}
\label{tab:privacy-targeted}
\resizebox{\textwidth}{!}{
\begin{tabular}{lccc ccccc ccc}
\toprule
\multirow{2}{*}{Target Attribute}
& \multicolumn{3}{c}{Distribution and Baselines}
& \multicolumn{5}{c}{Random-Split Telemetry Probe}
& \multicolumn{3}{c}{LOSO Telemetry Probe} \\
\cmidrule(lr){2-4}
\cmidrule(lr){5-9}
\cmidrule(lr){10-12}
& Pos. Rate & All-Positive F1 & Scenario-only F1
& AUROC & AP & Probe F1 & Prec.@$p{\ge}0.9$ & Cov.@$p{\ge}0.9$
& AUROC & Prec.@$p{\ge}0.9$ & Cov.@$p{\ge}0.9$ \\
\midrule
acute/recent
& 0.708 & 0.829 & 0.880
& 0.710 & 0.802 & 0.845 & 0.818 & 0.688
& 0.429 & 0.648 & 0.389 \\
high severity
& 0.704 & 0.826 & 0.800
& 0.765 & 0.869 & 0.800 & 0.960 & 0.463
& 0.448 & 0.745 & 0.611 \\
oncology-related
& 0.308 & 0.471 & 0.000
& 0.725 & 0.523 & 0.578 & 0.500 & 0.128
& 0.746 & 0.516 & 0.199 \\
\bottomrule
\end{tabular}
}
\end{table*}

Following the experimental setup described in Section~\ref{sec:eval_setup}, Table~\ref{tab:jailbreak-main} reports the jailbreak-wrapper transfer results under the mixed-positive setting. Since AutoDAN~\cite{Liu2023AutoDANGS} is included in the source-domain risk-positive pool, it is treated as a seen wrapper, while PAIR~\cite{chao2025jailbreakingPAIR} and CodeChameleon~\cite{Lv2024CodeChameleonPE} are used to evaluate transfer to unseen wrappers.

Overall, \textsc{RouteScan} demonstrates effective cross-wrapper transfer on both Qwen and DeepSeek across the two hardware configurations. 
The seen AutoDAN wrapper generally achieves the best or tied-best results, and the unseen PAIR and CodeChameleon wrappers retain high \(\mathrm{AUROC}\) and strong fixed-threshold performance. 
In particular, when we use AdvBench + AutoDAN as the source domain, DeepSeek transfers consistently to both unseen wrappers on A6000 and 5880. Identical \(\mathrm{F1}@0.5\) and \(\mathrm{Acc}@0.5\) values across some wrappers arise because they share the same benign set and produce identical false-positive decisions at the default threshold, rather than because duplicated jailbreak samples are used.

Transfer from StrongReject + AutoDAN to PAIR is more challenging. 
On DeepSeek/A6000, PAIR achieves an \(\mathrm{AUROC}\) of \(0.954\), but \(\mathrm{F1}@0.5\) and \(\mathrm{Acc}@0.5\) decrease to \(0.8480\) and \(0.8468\), respectively. 
This divergence suggests that the telemetry representation preserves effective cross-wrapper ranking information, whereas source-calibrated thresholds adapt less robustly to specific target distributions.

\subsection{Privacy Boundary Evaluation Results}
\label{sec:eval_privacy}

Table~\ref{tab:privacy-inversion} shows that the end-to-end inversion attacker has limited reconstruction capability. 
On the test split, Token F1 is 0.1601, with Jaccard and ROUGE-L at 0.1116 and 0.1194, respectively, indicating that the generated text has only limited lexical overlap with the original prompt. 
More importantly, all exact sensitive-field recovery metrics are 0. 
Thus, although the generated text may occasionally contain generic medical vocabulary or scenario-related descriptions, it does not stably recover the exact name, ID, date, age, clinical value, or condition field from the corresponding gold prompt.

We further use an LLM-based semantic audit to check whether rule-based extraction misses paraphrased or semantically equivalent leakage. 
Among 54 audited generations, the LLM judge finds no instance where the generated text recovers an exact sensitive fact from the corresponding gold prompt. 
These results suggest that, under the evaluated telemetry-to-text attacker, aggregated prefilling-phase GPU execution telemetry does not support reliable exact prompt reconstruction or exact sensitive-field recovery.

Table~\ref{tab:privacy-targeted} reports targeted coarse-attribute probing results. 
For acute/recent and high severity, random-split probe performance appears high, but the baselines indicate that much of this predictability is associated with label imbalance and scenario templates. 
Both attributes have positive rates around 0.70, and the all-positive and scenario-only baselines already achieve high F1 scores. 
Under leave-one-scenario-out (LOSO) evaluation, their \(\mathrm{AUROC}\) drops below 0.45, suggesting that the random-split performance does not correspond to a stable cross-scenario telemetry signal.

The oncology-related attribute shows a more nuanced pattern. 
Unlike the previous two attributes, oncology-related has a lower positive rate, a weaker all-positive baseline, and a zero scenario-only F1. 
This indicates that, under our coarse scenario-only baseline, the attribute is not directly predictable from scenario labels in the same way as acute/recent or high severity.
Its LOSO \(\mathrm{AUROC}\) remains at 0.746, indicating that condition-family information may leave a measurable but limited ranking signal in the aggregated GPU execution telemetry. 
However, this signal remains limited in operational terms: the high-confidence precision is only 0.500 in the random split and 0.516 under LOSO, with low coverage in both cases. 
Consequently, the oncology-related results serve better as an exceptional indicator for privacy boundaries: while aggregated GPU telemetry is insufficient to support exact prompt reconstruction or high-fidelity sensitive field recovery, it may still preserve rankable, weak profiling traces for certain coarse-grained medical attributes.

In summary, prefilling-phase GPU telemetry resists exact prompt and sensitive-field reconstruction. While coarse-grained profiling carries residual inference risk, our LOSO analysis demonstrates that apparent predictability is often a byproduct of dataset bias rather than stable telemetry leakage, supporting a bounded empirical privacy guarantee.

\section{Conclusion}
We propose \textsc{RouteScan}, a harmful-behavior auditing system that identifies potentially harmful LLM interactions from hardware-derived expert routing telemetry observed in prefilling-phase GPU execution. It constructs request-level representations from expert-execution fingerprints and selects transferable features through hybrid scoring and entropy-based adaptive support, enabling robust risk detection without exposing plaintext prompts, model outputs, or internal model states. Across open-source MoE models, \textsc{RouteScan} achieves AUROC above 0.91 on diverse harmful domains and unseen jailbreak wrappers. Empirical evaluations further show that the aggregated telemetry resists plaintext prompt inversion and exact sensitive-field recovery under the evaluated attacks. Our work provides a new paradigm for AI governance by demonstrating the feasibility of GPU telemetry-based auditing with limited exposure to user information.

\bibliographystyle{IEEEtran}
\bibliography{sample-base}

\end{document}